\documentclass[onecollarge]{svjour2}       
\smartqed  
\usepackage{graphicx}
\begin{document}

\title{Stationary waves in a supersonic flow of a two-component Bose gas
}


\author{L.Yu. Kravchenko         \and
        D.V. Fil 
}


\institute{D.V. Fil \at
              Institute for Single Crystals, National Academy of
Science of Ukraine, Lenin av. 60, Kharkov 61001, Ukraine \\
              Tel.: +38-057-3410492\\
              Fax: +38-057-3409343\\
              \email{fil@isc.kharkov.ua}           
           \and
           L.Yu. Kravchenko \at
              Institute for Single Crystals, National Academy of
Science of Ukraine, Lenin av. 60, Kharkov 61001, Ukraine }

\date{Received: date / Accepted: date}

\maketitle

\begin{abstract}
A stationary wave pattern occurring in a flow of a two-component
Bose-Einstein condensate past an obstacle is studied. We consider
the general case of unequal velocities of two superfluid
components. The Landau criterium applied to the two-component
system determines a certain region in the velocity space in which
superfluidity may take place. Stationary waves arise out of this
region, but under the additional condition that the relative
velocity of the components does not exceed some critical value.
Under increase of the relative velocity the spectrum of the
excitations becomes complex valued and the stationary wave pattern
is broken. In case of equal velocities two sets of stationary
waves that correspond to the lower and the upper Bogolyubov mode
can arise. If one component flows and the other is at rest only
one set of waves may emerge. Two or even three interfere sets of
waves may arise if the velocities approximately of equal value and
the angle between the velocities is close to $\pi/2$. In two
latter cases the stationary waves correspond to the lower mode and
the densities of the components oscillate out-of-phase. The ratio
of amplitudes of the components in the stationary waves is
computed. This quantity depends on the relative velocity, is
different for different sets of waves, and varies along the crests
of the waves. For the cases where two or three waves interfere the
density images are obtained.
\end{abstract}

\section{Introduction}
\label{intro} A unique feature of two-component Bose-Einstein
condensates (BEC) is the possibility for two superfluids to flow
with different velocities. The properties of such systems can be
described by the three-velocity hydrodynamics (one normal and two
superfluid velocities). This feature was already noticed by
Khalatnikov \cite{hal}. The modern three-velocity superfluid
hydrodynamic theory was formulated in the paper by Andreev and
Bashkin \cite{ab}. As was shown in \cite{ab} the specifics of the
three-velocity superfluid hydrodynamics  is the presence of a
non-dissipative drag between the components. The drag effect
emerges at nonzero relative velocity of the components and
consists in a dependence of the superfluid current of one
component on the gradient of the phase of the order parameter of
the other component. The microscopic theory of the non-dissipative
drag effect was developed in  \cite{fil1,fil}. An important
question that arises in
the three-velocity
superfluid hydrodynamics is the question on critical velocities.
At equal velocities of the components the answer to this question
can be obtained from the Galilean transformation. It yields that
under neglecting of vortex excitations the critical velocity
coincides with the minimal phase velocity of the lower Bogolyubov
mode. In a wider sense one can introduce two critical velocities
$c_-$ and $c_+$, one is for the lower mode and the other is for
the upper mode \cite{nb,bi2}.

In case of unequal velocities  of the components the situation
becomes more complicated. The question
was considered in Refs. \cite{yukalov,6}. The authors of
\cite{yukalov} have obtained the dispersion equation for the
spectrum of elementary excitations in the presence of superfluid
flows. However the analysis of critical velocities in
\cite{yukalov} was based on the implication that the Landau
criterium can be formulated as the condition on the relative
velocity of the superfluid components. Such an implication can be
put in question since there are two independent relative
velocities in the three-velocity theory. In \cite{6} the Landau
criterium was formulated as the condition of positiveness of
energies of elementary excitations in a reference frame connected
with a normal component. It yields a joint condition on absolute
values of the superfluid velocities of the components and on the
angle between their directions. If one component is at rest the
superfluidity condition \cite{6} is reduced to \cite{yukalov}.

The analysis carried out in \cite{6} shows that the Landau
criterium may be fulfilled if one or even both superfluid
velocities exceed the velocity of the lower mode $c_-$(in the
latter case the velocities should have different directions). In
view of unusual behaviour of critical velocities in such systems
it is interesting to consider how this behavior can reveal itself
in experiments. Two-component BECs have been realized
experimentally in ultracold alkali metals gases confined in
magnetic and magneto-optical traps. Two components may correspond
to different hyperfine Zeeman states of the same isotope
\cite{Rb2,Rb-n}, or to different isotopes \cite{K-Rb,K-Rb1,Rb3}.
One of the methods to determine critical velocities for trapped
ultracold gases consists in the observation of density excitations
induced by some object moving through the condensate \cite{exp}
(usually a laser beam is used as such an object). A motion of an
object in a two-component gas with nonzero relative velocity of
the superfluid components corresponds to the general case of the
three-velocity hydrodynamics.

The Bogolyubov spectrum has a dispersion. Therefore, a motion of
an object through a superfluid system (or a superfluid flow past
an obstacle) can lead to an occurrence of stationary waves (the
waves whose crests remain at rest relative to the obstacle). Such
an effect called "ship waves" is well-known \cite{wi}. It was
considered by Kelvin for the waves generated on a water surface by
a ship moving in a deep water. Stationary waves in a one-component
quasi-two-dimensional BEC were studied in \cite{ship}. It was
shown that in a superfluid that flows past a point obstacle (the
obstacle size is less than the healing length) the stationary wave
pattern is similar to one for stationary capillary waves. The
effect takes place if the superfluid velocity $s$ exceeds the
minimal phase velocity of the Bogolyubov mode $c_0$. The
stationary waves arise outside the Mach cone bounded by arms
directed at the angles $\vartheta =\pm \arcsin(c_0/s)$ relative to
the flow.

In recent papers the solitons \cite{bi2,16} and stationary waves
\cite{16} induced by an obstacle in a two-component superfluid
system were studied (the paper \cite{16} was published as the
electronic preprint when the present study was almost completed).
But the authors of \cite{bi2,16} considered only the case of equal
superfluid velocities (a relative velocity of the components is
equal to zero).

In the present paper we put emphasis on the general case of
nonzero relative velocity. In Sec. \ref{sec2} we obtain the
equation for the spectrum and the eigenvectors of collective
excitations and define two more critical velocities (in addition
to $c_-$ and $c_+$). One of them is the maximum critical velocity
$c_m$ for a given component ($c_m>c_-$). The superflow at the
velocity $s\to c_m$ can be reached if the other component is at
rest. The other is the relative critical velocity $c_{\rm sep}$
($c_{\rm sep}>c_m$). If the relative velocity exceeds $c_{\rm
sep}$, the frequency of the lower mode becomes complex valued. The
latter signals for an instability of the two-component system with
respect to a spatial separation of the components. In
Sec.\ref{sec3} the equation that describes the stationary wave
pattern is obtained. It is shown that the stationary waves emerge
if the Landau criterium of superfluidity is violated (the energies
of the excitations with certain wave vectors becomes negative),
but the system remains stable with respect to the spatial
separation (the critical relative velocity is not achieved). A
number of stationary wave patterns are presented. It is shown that
depending on the velocities of the components several
qualitatively different situations are possible. If the velocities
are the same in modulus and in direction, one set of stationary
waves appears at $s>c_-$, and another set adds  at $s>c_+$. The
phase separation does not occur. If only one component moves with
the velocity $s$, one set of stationary waves is formed at $s>c_m$
but if $s$ reaches $c_{\rm sep}$  the phase separation occurs. If
the velocities are equal in modulus and the angle between their
directions is $\pi/2$ or close to $\pi/2$ two or three sets of
stationary waves occur at $s_1=s_2>c_m$. In Sec. \ref{sec4} we
investigate the structure of the stationary waves. The densities
of the components always oscillate out-of-phase in the waves that
correspond to the lower mode . The ratio of  the amplitudes of the
oscillations of the components depends on the relative velocity
and varies along the crests of the waves. For complex density
patterns where two or three waves interfere the density plots  are
presented.  It is established that the stationary waves are
visible in total density images as well as relative density
images, but in most cases relative density images are more
contrast. The only exception is the stationary waves that are
exited at $s>c_+$ and correspond to the upper Bogolyubov mode.

\section{The spectrum of collective modes}\label{sec2}

To analyze the stationary waves in a two-component superfluid
system one should obtain the collective modes spectrum in a moving
condensate. It can be found from the matrix version of the
Gross-Pitaevskii equation
\begin{eqnarray}\label{19}
i \hbar \frac{\partial \psi_1}{\partial t} = - \frac{\hbar^2}{2
m_1} \triangle \psi_1 + \gamma_1 |\psi_1|^2 \psi_1 + \gamma_{12}
|\psi_2|^2 \psi_1, \cr
 i \hbar \frac{\partial \psi_2}{\partial t} =
- \frac{\hbar^2}{2 m_2} \triangle \psi_2 + \gamma_2 |\psi_2|^2
\psi_2 + \gamma_{12} |\psi_1|^2 \psi_2,
\end{eqnarray}
where $\psi_i$ are the wave functions of the components, $m_i$ are
the masses of the particles,
\begin{equation}\label{47}
\gamma_i = \frac{4 \pi \hbar^2 a_i}{m_i}, \; \gamma_{12} = \frac{2
\pi \hbar^2 (m_1 + m_2) a_{12}}{m_1 m_2}
\end{equation}
are the interaction constants ($a_i$ and $a_{12}$ are scattering
lengthes).

Here we restrict the consideration by the most convenient for the
analysis symmetric case for which the components have equal masses
of the particles $m_1 = m_2 = m$, equal densities $n_1=n_2=n_0$ of
the components and equal interaction constants $\gamma_1 =
\gamma_2 = \gamma$. While this case is quite specific, it is
possible to produce equal $\gamma_i$ using a Feshbach resonance.
The symmetric case may also correspond to two quasi
two-dimensional Bose clouds with a strong dipole interaction
separated by a rather high (but thin) barrier that suppress the
tunneling  (see \cite{fil1}). On the qualitative level the results
obtained for the symmetric case hold for the general case where
such a symmetry between the components is broken. We assume the
interaction between the particles of the same component is
repulsive ($\gamma> 0$), and the stability condition with respect
to  a spatial separation of the components $(\gamma>
|\gamma_{12}|)$ is fulfilled. Going ahead we note that in case of
different velocities that condition is necessary but not
sufficient one.

If the temperature is much less than the temperature of the
Bose-Einstein condensation the wave functions of the components can
be presented as a sum of a large stationary part and a small
fluctuating part
\begin{equation}\label{30}
\psi_i (\mathbf{r},t) = \psi_{0i}(\mathbf{r},t) + \delta \psi_i
(\mathbf{r},t).
\end{equation}
The stationary part of the condensate wave function reads as
\begin{eqnarray}\label{31}
\psi_{0i}(\mathbf{r},t) =  \sqrt{n_0} \,e^{i
\varphi_i(\mathbf{r})}e^{-\frac{i \mu_i t}{\hbar}} ,
\end{eqnarray}
where $\mu_i = \displaystyle \frac{m_i \mathbf{s}_i^2}{2} +
(\gamma + \gamma_{12}) n_0$ are the chemical potentials of the
components, and $\mathbf{s}_j = \displaystyle
\frac{\hbar}{m}\nabla \varphi_j$ are their superfluid velocities.
We will search for the fluctuating part of Eq. (\ref{30}) in the
form:
\begin{eqnarray}\label{31a}
\delta \psi_i (\mathbf{r},t) = e^{i \varphi_i(\mathbf{r})} e^{-
\frac{i \mu_i t}{\hbar}} \left[\textrm{u}_i(\mathbf{r}) e^{-i
\omega t} + \textrm{v}_i^*(\mathbf{r}) e^{i\omega^* t}\right].
\end{eqnarray}
Here the functions $\textrm{u}_i$ and $\textrm{v}_i$ are the plane
waves
\begin{equation}\label{33}
\textrm{u}_i(\mathbf{r}) = A_i e^{i \mathbf{k} \mathbf{r}}, \;
\textrm{v}_i(\mathbf{r}) = B_i e^{i \mathbf{k} \mathbf{r}}.
\end{equation}
Substituting Eqs. (\ref{30}) -- (\ref{33}) into Eq. (\ref{19}), in
a linear in fluctuations approximation we obtain the following
equation for the excitation energies $\hbar\omega$ and the
eigenvectors:
\begin{equation}\label{40} \mathbf{M} \mathbf{V}= \omega
\mathbf{V},
\end{equation}
where
\begin{equation}\label{41}
\mathbf{M} = \left(%
\begin{array}{cccc} \displaystyle
\left(\frac{k^2}{2} + 1 + \mathbf{s}_1 \mathbf{k}\right) & 1 & \gamma' & \gamma' \\
-1 &\displaystyle -\left(\frac{k^2}{2} + 1 - \mathbf{s}_1 \mathbf{k}\right) & -\gamma' & -\gamma' \\
\gamma' & \gamma' &\displaystyle  \left(\frac{k^2}{2} + 1 + \mathbf{s}_2 \mathbf{k}\right) & 1 \\
-\gamma' & -\gamma' & -1 &\displaystyle  -\left(\frac{k^2}{2} + 1 - \mathbf{s}_2 \mathbf{k}\right)\\
\end{array}%
\right)
\end{equation}
and
\begin{equation}\label{ev}
    \mathbf{V}=\left(%
\begin{array}{c}
  A_1 \\
  B_1 \\
  A_2 \\
  B_2 \\
\end{array}%
\right)
\end{equation}
 Here and below all physical quantities are expressed in terms of
dimensionless length and time
\begin{eqnarray}\label{22}
\tilde{r_i} = \frac{r_i}{\sqrt{2} \xi},\;  \tilde{t} =
\frac{c_0}{\sqrt{2} \xi} t,
\end{eqnarray}
where $c_0 =\sqrt{\gamma n_0/m}$ is a sound velocity in a
one-component condensate, and $\xi =\hbar/\sqrt{2 m \gamma n_0}$
is a healing length.  We also define the dimensionless parameter
of the interspecie interaction $\gamma' = \gamma_{12}/\gamma$.

The $u-v$ transformation procedure (\ref{31a}) is equivalent to
the diagonalization of a quadratic form on Bose operators
\cite{pit,bogol}. The matrix $\mathbf{M}$ has four eigenvalues.
The components of the eigenvectors of Eq. (\ref{40}) can be
normalized as $|A_1|^2-|B_1|^2+|A_2|^2-|B_2|^2=\pm 1/V$ ($V$ is
the volume of the system). For two physical modes the norm of the
eigenvector should be positive \cite{bogol}. The spectra of the
physical modes read as
\begin{equation}\label{42}
\omega_{\pm} = \frac{(\mathbf{s}_1+\mathbf{s}_2) \mathbf{k} 
}{2} + \sqrt{k^2 \left(1 + \frac{k^2}{4}\right) +
\frac{(\mathbf{s}_- \mathbf{k} )^2}{4} \pm \sqrt{k^2\left(1+
\frac{k^2}{4}\right)(\mathbf{s}_- \mathbf{k} )^2 + k^4
\gamma'^2}},
\end{equation}
where $\mathbf{s}_- = \mathbf{s}_1 - \mathbf{s}_2$ is the relative
superfluid velocity.
 At $s_1=s_2 = 0$ the spectrums (\ref{42}) has
the standard Bogolyubov form
\begin{equation}\label{42-1}
\omega_{\pm}=k\sqrt{1\pm |\gamma'|+k^2/4}.
\end{equation}
As is clear from Eq. (\ref{42-1}), the stability condition with
respect to a spatial separation $\gamma> |\gamma_{12}|$ is the
requirement for  the excitation spectrum be real valued. At small
$\mathbf{k}$ the excitation spectrum is a sound one and the
velocities of the modes are equal to $c_\pm = \sqrt{1\pm
|\gamma'|}$. The Landau criterium  requires  the spectrum
(\ref{42}) be positive valued  at all wave vectors. One can see
from Eq.(\ref{42}) that at $\mathbf{s}_1 = \mathbf{s}_2
=\mathbf{s}$ the Landau criterium is reduced to the inequality $s
<c_-$. If only one component flows ($s_2=0$), the Landau criterium
requires a fulfilment of the inequality $s_1 <c_m=c_-\sqrt{1 +
|\gamma'|} $. If the velocity of a given component exceeds $c_m$,
the superfluidity condition is broken irrespective of a value and
direction of the velocity of the other component \cite{6}, i.e.
$c_m$ can be called the maximum critical velocity.

At the velocities for which the energy (\ref{42}) is negative
valued in some range of ${\bf k}$ the Landau criterium is broken
and stationary waves can occur in the system. In contrast to a
one-component system, in a two-component system an increase of
superfluid velocities may result in that the spectrum of the lower
mode be complex valued. As follows from Eq. (\ref{42}), the
spectrum remains real valued if the relative velocity satisfies
the condition
\begin{equation}\label{17}
|\mathbf{s}_1 - \mathbf{s}_2|< c_{\rm sep}= 2 c_- .
\end{equation}
At complex valued  frequencies (\ref{42})  the amplitude of
excitations grows with time that leads to a destruction of a
homogeneous state and to a spatial separation of the components
(or stratification). In contrast to Ref. \cite{yukalov}, we
consider the condition of stratification and the Landau criterium
as different conditions. The Landau criterium yields a joint
restriction on both superfluid velocities (and its mutual
direction), while the condition of stratification is a restriction
only on the modulus of the relative velocity. Note that $c_{\rm
sep}> c_m $, and under increase of the velocity of a given
component, first, the Landau criterium is violated, and then,
after further increase, the stability condition with respect to
the stratification is broken. One can show, that the same
situation takes place under simultaneous increase of two
velocities (at nonzero relative velocity). Only at equal in value
and oppositely directed superfluid velocities the Landau criterium
and the stability condition are broken at the same point.

Since we are interested in stationary waves in a homogeneous (not
stratified) system, we will consider only the velocities for which
the condition (\ref{17}) is satisfied.

\section{The stationary wave pattern}\label{sec3}

Let us consider a two-component BEC that flows past an obstacle
situated at the origin of coordinates. We assume the system is
quasi-two-dimensional, i.e. it is thin enough  to neglect the
dependence of the condensate wave function on the transverse
coordinate, and to consider all vector quantities as
two-dimensional ones. If the size of an obstacle is much less than
$ \xi $ it can be considered as a point one. Under violation of
the Landau criterium the obstacle behaves as a point source of
waves (below we are only interested in stationary waves). The
waves propagate from an obstacle with the group velocities
$\mathbf{v}_{g\pm} =
\partial \omega_{\pm} / \partial \mathbf{k}$, where $\omega_\pm =\omega_\pm (k_x, k_y)$
are given by Eq. (\ref{42}) at $k_z=0$. Here and below the index
$+(-)$ corresponds to the waves generated by the upper (lower)
mode. The direction of  the propagation for the wave with a given
$\mathbf{k}$ is defined by the expression
\begin{equation}\label{9}
\tan\chi_{\pm} =\frac{\partial \omega_{\pm}/\partial k_y}{\partial
\omega_{\pm}/\partial k_x},
\end{equation}
where $\chi_+ (\chi_-)$ is the angle between the group velocity
direction for a given mode and the axis $x$.

For the stationary wave the frequency (\ref{42}) is equal to zero,
and the components of the wave vector  $ \mathbf{k} = (k_x, k_y) $
are related by the equation
\begin{equation}\label{60}
     \omega_\pm(k_x, k_y)=0.
\end{equation}
It is convenient to use the angle $\eta$ between the wave vector and
the opposite direction of the axis $x$ as an independent parameter,
i.e.
\begin{equation}\label{12-0}
 \mathbf{k} = (- k \cos\eta, k \sin\eta).
\end{equation}
Let us denote the angle between the velocities by $\theta $ and
select the axis $x$ along the bisectrix of this angle ($
\mathbf{s}_1 = (s_1 \cos \frac{\theta}{2}, - s_1 \sin
\frac{\theta}{2})$ and $\mathbf{s}_2 = (s_2 \cos \frac{\theta}{2},
s_2 \sin \frac{\theta}{2})$). For the stationary waves Eq.
($\ref{60}$) yields the following dependence of the wave number
$k$ on the angle $\eta$
\begin{equation}\label{12}
\begin{array}{c}
  k_{\pm}(\eta) = \sqrt{2} \biggl[-2 + s_1^2 \cos^2\left(\eta -
\frac{\theta}{2}\right) + s_2^2 \cos^2\left(\eta+
\frac{\theta}{2}\right) \mp \\ \left.
  \sqrt{4 \gamma'^2 + \left(s_1^2
\cos^2\left(\eta - \frac{\theta}{2}\right) - s_2^2
\cos^2\left(\eta
+ \frac{\theta}{2}\right)\right)^2}\right]^{1/2}. \\
\end{array}
\end{equation}

\begin{figure}
  \includegraphics[width=6cm]{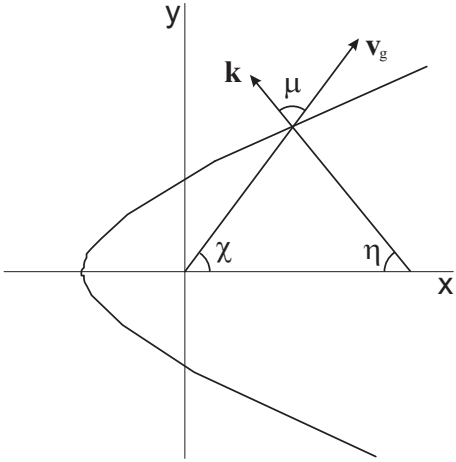}
\caption{The scheme of a wavefront crest} \label{scheme}
\end{figure}

A wave crest line is a line of a constant phase. For the
stationary wave the phase can be obtained from the equation
\begin{equation}\label{13}
\phi(\mathbf{r}) = \int_0^{\mathbf{r}} \mathbf{k} d\mathbf{r}.
\end{equation}
In Eq. (\ref{13}) the integral is  along a straight line going out
from the origin of coordinates and directed parallel to the group
velocity $\mathbf{v}_{g\pm}$. The angle $\mu$ between $\bf{k}$ and
$ \bf{r}$ is defined by the expression
\begin{equation}\label{14-0}
    \mu = \pi - \eta - \chi_\pm
\end{equation}
(see Fig. \ref{scheme}).

According to Eq.(\ref{13}), the quantity $r$ for the points at the
wave crest with a given phase $\phi$ satisfies the equation
\begin{equation}\label{15}
r=\frac{\phi}{k \cos\mu}.
\end{equation}
Substituting Eq. (\ref{14-0}) into Eq. (\ref{15}), we get the
equations for the stationary wave crest coordinates $x=r \cos\chi$
and $y=r \sin\chi$ in a parametric form
\begin{eqnarray}\label{16}
x_{\pm}= - \frac{\phi}{k_{\pm}(\eta)\cos\eta \left[1 -
\tan\chi_{\pm}\tan\eta\right]}, \cr y_{\pm}= -
\frac{\phi\tan\chi_{\pm}}{k_{\pm}(\eta)\cos\eta \left[1 -
\tan\chi_{\pm} \tan\eta\right]}.
\end{eqnarray}
In Eq. (\ref{16}) the values $\tan\chi_{\pm}$ are the functions of
the parameter $\eta$. The explicit form of these functions is
given by Eq. (\ref{9}), in which after differentiation one should
substitute Eqs. (\ref{12-0}) and (\ref{12}). The range of values
of $ \eta $ is determined by the condition for the function $k_\pm
(\eta) $ be real valued. Eqs. (\ref{16}) allow to draw the
stationary wave pattern for arbitrary values of $s_1$, $s_2$ and $
\theta $.

Let us consider some special cases. For definiteness, we choose
the parameter $\gamma'=0.5$. Such a choice corresponds to $c_- =
0.707$, $c_m=0.866$ and $c_+ = 1.22$ (in $c_0$ units).

\begin{figure}
  \includegraphics[width=6cm]{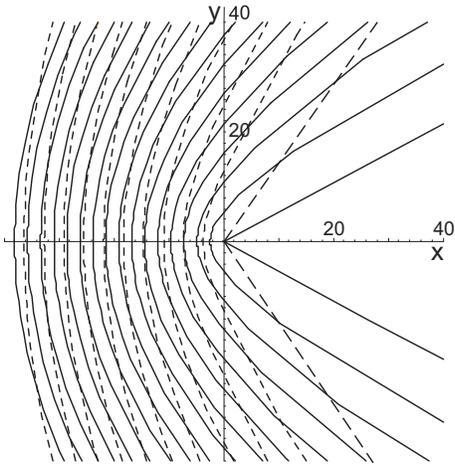}
\caption{The wave pattern forming at the flow of components with
equal velocities $(s=1.5)$. The waves caused by the lower and the
upper modes are shown with solid and dashed lines, respectively.}
\label{waves10}
\end{figure}

1. At equal superfluid velocities $\mathbf{s}_1 = \mathbf{s}_2
=\mathbf{s}$ Eqs. (\ref{16}) yield the expected result. At $s>
c_-$ a set of stationary waves corresponding to the lower mode
arises. The crests for these waves are outside the cone bounded by
the arms $\chi_1 =\pm \arcsin(c_-/s) $. If the velocities $s> c_+$
are reached the second set of stationary waves appears. It
corresponds to the upper mode and situated outside the arms $\chi
=\pm \arcsin (c_+/s) $. Since $s_- = 0$, the stratification does
not occur at any $s$. As an example, the stationary wave pattern
for $s=1.5$ is shown in Fig.\ref{waves10}.

2. If only one component (say, the component 1) flows, the
stationary waves arise at the velocities $s_1> c_m $. These waves
correspond to the lower mode. The waves are outside the cone $
\chi_- =\pm \arcsin(c_m/s) $. In this case the frequency of the
upper mode is always positive and the second set of waves cannot
arise. At $s_1> c_{\rm sep}$ (for the  parameters chosen $c_{\rm
sep} =1.41$) the system becomes unstable with respect to the
stratification.  The stationary wave pattern for the flow of one
component with $s_1=1.0$ is shown in  Fig.\ref{waves1}.

3. It is interesting to analyze the situation when the angle
between the velocities is  $\theta=\pi/2$. In this case the
obstacle emits waves only when the velocity of at least one
component exceeds $c_m$. Let us consider a more specific case of
the velocities equal in magnitude $s_1=s_2=s$. Then the velocity
range in which the stationary waves occur is limited by the
condition $c_m <s <\sqrt{2} c_- $ (for the parameters chosen
$0.866 <s <1$). In Fig.\ref{waves4} we present the stationary wave
pattern for $s=0.9$. One can see that in such a situation two sets
of stationary waves arise. It is important to emphasize that both
sets correspond to the lower mode (the frequency of the upper mode
remains positive for the velocities in the range $c_m <s <\sqrt{2}
c_-$).

In the limit $ \gamma'\to 0$ (that corresponds to the absence of
the interaction between the components) each component should have
its own set of stationary waves at any relative directions of the
velocities. For rather large $\gamma'$ this feature survives only
in a close vicinity to $\theta=\pi/2$. There should be a smooth
transition from $\theta =\pi/2$ to $\theta=0$. The analysis of the
wave patterns at different $\theta$ shows that at intermediate
$\theta$ a quite complicated pattern emerges: the crests of a
given set end with cusps, and bridges connect them with crests
from the other set. Under decrease of $\theta$ the bridges and
cusps disappear and the wave pattern becomes similar to one for $
\theta=0$. The stationary wave pattern with cusps and bridges is
shown in Fig. \ref{waves11}.

In general, the stationary wave pattern is qualitatively similar
to one of presented in  Figs. \ref{waves10}-\ref{waves11}.

\section{The density pattern for the stationary waves}\label{sec4}

There is a number of methods of probing  BECs to get their density
profiles (see \cite{ket}). Advanced technics was developed for the
study of  density profiles of spinor BECs \cite{c0,c1,c2,c3,prl05}
(that in certain sence can be considered as two-component ones).
In particular, the density and the spin-density profiles of an
atomic cloud were measured with a high resolution by the
polarization-dependent phase-contrast imaging method
\cite{prl05,sk07}.

In view of modern experimental possibilities it is important to
find the ratio of the components in the stationary waves and to
determine specific features of the total density and the relative
density patterns for the stationary wave in two-component systems.

Since the stationary wave is the eigenmode, the ratio of the total
density and relative density amplitudes can be found from the
corresponding eigenvector (\ref{ev}).

The density of a given component is the square modulus of the
condensate wave function $ \rho_i = | \psi_i |^2$. This quantity
can be presented as a sum of the unperturbed density $n_0$ and the
perturbation $\delta n_{i}$ caused by the stationary wave. Using
the equation for the condensate wave function (\ref{30}) and Eqs.
(\ref{31}) -- (\ref{33}), we obtain
\begin{equation}\label{38} \delta n_{i}
= \psi_{0} \delta \psi_{i}^* + \delta \psi_{i} \psi_{0}^*=\delta
\rho_{i} \cos(\mathbf{k} \mathbf{r}),
\end{equation}
where
\begin{eqnarray}\label{46}
\delta\rho_{i} = {\cal A} \sqrt{n_0}  (A_{i} + B_{i}).
\end{eqnarray}
here ${\cal A} $ is the amplitude of a given eigenmode. The
amplitude ${\cal A}$ depends on the intensity of the source
(obstacle) and on the distance from the source. As follows from
Eqs. (\ref{40},\ref{41}), the components of the eigenvectors are
real valued quantities. That is why Eq. (\ref{46}) may correspond
to in-phase oscillations of the densities  or to the oscillation
with the phase shift equal to $\pi$ (out-of-phase). Here we
consider the case $\gamma'>0$ for which the lower Bogolyubov mode
corresponds to out-of-phase oscillations and the higher mode - to
in-phase oscillations.

The total density and the relative density oscillation amplitudes
in the stationary wave are $\delta\rho_\pm=\rho_1\pm \rho_2$. The
specific of the symmetric case (components with equal masses,
densities and interaction constants) is that in the absence of the
flow the oscillations of the total density vanish for the lower
Bogolyubov mode ($\delta\rho_+=0$), and the oscillations of the
relative density are absent in the higher mode ($\delta\rho_-=0$).
As was shown in the previous section, in most cases the stationary
waves are caused by the lower mode. Therefore it is important to
clarify whether is the total density disturbed in the stationary
waves.

Using the analytical expressions for the eigenvectors of the
matrix (\ref{41}) one finds that the exact relations $\delta\rho_1
=-\delta\rho_2$ for the lower mode and $\delta\rho_1 =
\delta\rho_2$ for the upper mode are hold in case of equal
velocities (${\bf s}_1={\bf s}_2$). In means that in the
stationary wave pattern shown in Fig. \ref{waves10} the lower mode
set is visible in the relative density image, while the upper mode
set - in the total density image.

\begin{figure}
    \includegraphics[width=6cm]{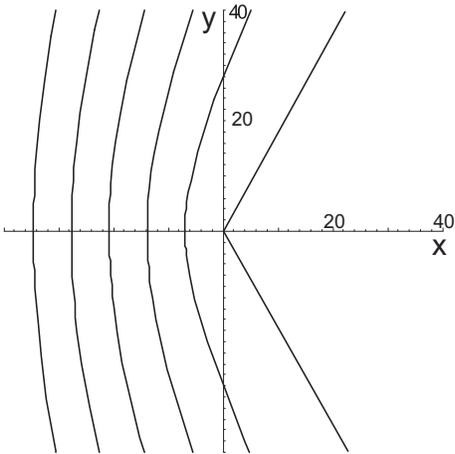}
\caption {The wave pattern forming at the flow of one component
with the velocity $s_1 = 1.0 \, (s_2 = 0)$. The axis $x$ is along
the flow direction.} \label{waves1}
\end{figure}

\begin{figure}
    \includegraphics[width=6cm]{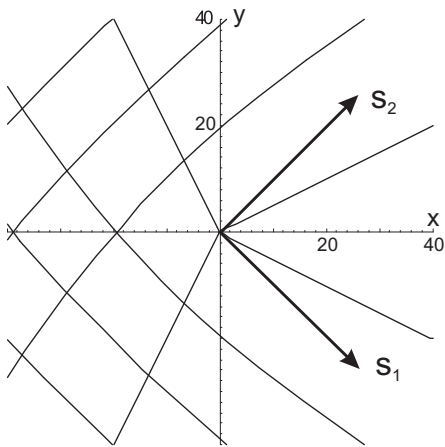}
\caption {The wave pattern forming by two orthogonally directed
flows with the velocities $s_1 = s_2 = 0.9$. The directions of the
velocities are shown by the arrows.}\label{waves4}
\end{figure}

\begin{figure}
    \includegraphics[width=6cm]{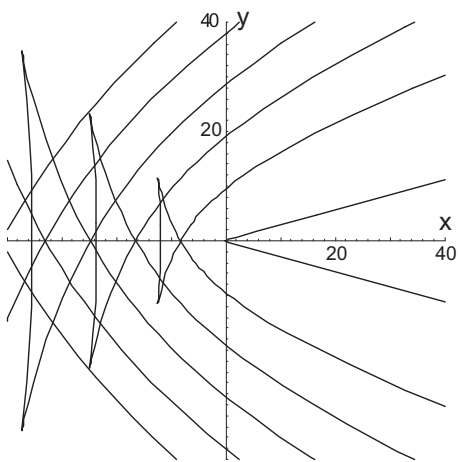}
\caption {The wave pattern  at $s_1=s_2 =1$ and $ \theta = 0.45 \pi
$}\label{waves11}
\end{figure}

\begin{figure}
    \includegraphics[width=8cm]{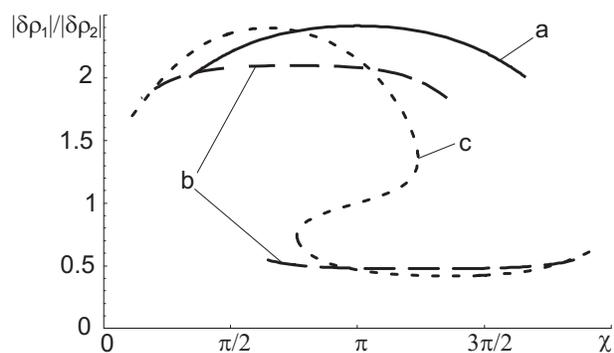}
\caption {The ratio of densities of the components in the
stationary waves along the chests for: a) one component flow $(s_1
= 1, s_2 = 0)$; b) two component flow in mutually orthogonal
directions
 $(\theta = \pi/2, s_1 = s_2 = 0.9)$; c) the case of $\theta = 0.45\pi, s_1 =
s_2 = 1$.} \label{density1}
\end{figure}

In the general case ${\bf s}_1\ne {\bf s}_2$ we find that
$\delta\rho_1 \ne -\delta\rho_2$ and the total density
oscillations are nonzero for the stationary waves that correspond
to the lower mode, in particular, for the pattern shown in Figs.
\ref{waves1} - \ref{waves11}. For these cases the ratios
$|\delta\rho_1|/|\delta\rho_2|$ along the crests are shown in Fig.
\ref{density1}. Since this ratio differs from unity, both,
$\delta\rho_+$ and $\delta \rho_-$ are nonzero and stationary wave
patterns should be visible in the total density image as well as
in the relative density image. We also would like to point out the
feature that follows from Fig. \ref{density1}. If only one
component flows past the obstacle the stationary waves contain
admixture of both components, but the flowing component has larger
amplitude.

In cases shown in Figs. \ref{waves1},\ref{waves4} two or three
sets of stationary waves interfere. Thus the density pattern
depends not only on the eigenvectors but on the amplitudes of the
eigenmodes, and another approach should be used to analyze the
density profiles. Here we use the approach developed in \cite{16}.

The interaction of the Bose gas with the obstacle is described by
the Hamiltonian $$H_{int}=\int d {\bf r} \sum_i V_i (\mathbf{r}) |
\psi_i |^2,$$ where $V_i (\mathbf{r})$ is the potential of the
interaction between the $i$-th component and the obstacle.
Considering the obstacle as a point source and assuming that the
obstacle interacts identically with both components we set $V_1
(\mathbf{r}) =V_2 (\mathbf{r}) =V_0 \delta({\bf r})$.
Respectively, the terms $V_0 \delta({\bf r}) \psi_i$ should be
added to the right hand parts of the Gross-Pitaevskii equations
(\ref{19}). The Gross-Pitaevskii equations can be rewritten in
terms of density and velocity fields (the hydrodynamic form).
Linearizing the hydrodynamic equations with respect to the density
and velocity fluctuations (see, for instance, \cite{6}), and
excluding the velocity fluctuations we arrive at the equations for
the Fourier components of $\delta n_i$:
\begin{equation}\label{111}
\left[- \left(\mathbf{s}_i \mathbf{k}\right)^2 + k^2 \left(1 +
\frac{k^2}{4}\right)\right] \delta n_{i}({\bf k}) + k^2 \gamma'
\delta {n}_{3-i}({\bf k}) = - k^2 \tilde{V}_0,
\end{equation}
where $\tilde{V}_0=V_0/\gamma $. In deriving Eq. (\ref{111}) we
take into account that all time derivatives are zero for the
stationary waves.

Solving Eq.(\ref{111}) and taking the inverse Fourier
transformation we get
\begin{equation}\label{111a}
\delta n_i = - \frac{4 V_0}{\pi^2\gamma\xi^2} {\rm Re}
\int_{-\pi/2}^{\pi/2} d\eta \int_0^{\infty} k d k \frac{ 1 - \gamma'
+ \frac{k^2}{4} - \frac{\left(\mathbf{s}_i \mathbf{k}\right)^2}{k^2}
}{(k^2 - k_+^2)(k^2 - k_-^2)} \, e^{i k r \cos \mu},
\end{equation}
where the quantities $k_\pm$ are given by Eq. (\ref{12}). The
angle $\mu(\eta)$ is determined by Eqs. (\ref{14-0}) and
(\ref{9}). The integral over $k$ in Eq.(\ref{111a}) can be
evaluated analytically with the use of the residue theorem. The
poles $k=k_-$ and $k=k_+$ (for real $k_\pm$) yield the
contribution of the lower and the upper Bogolyubov modes,
respectively. The integral over $\eta$ is evaluated by the
stationary phase method. The number of stationary phase points
coincides with the number interfered waves in the stationary wave
pattern (1 point for Figs. \ref{waves10} and \ref {waves1}, 2
points for Fig. \ref{waves4} and 3 points for Fig. \ref{waves11}).

The  answer  can be presented in the following form
\begin{equation}\label{112}
 \delta n_i\approx \frac{V_0}{\gamma\xi^2}\sum_{i,\lambda}
C_{i,\lambda}(\chi)\frac{\cos\left [\kappa_{i,\lambda}(\chi)
r-\pi/4\right]}{\sqrt{r}},
\end{equation}
where the index $\lambda$ numbers the sets that contribute to the
stationary wave pattern in a given sector of $\chi$. In Eq.
(\ref{112}) short range terms $\propto 1/r^2$ are omitted. The
coefficients $C_{i,\lambda}(\chi)$ and $\kappa_{i,\lambda}(\chi)$
are quite complicate and we do not present  the explicit
expressions here.

The case ${\bf s}_1={\bf s}_2={\bf s}$ can be analyzed directly
from Eq. (\ref{111a}). At such velocities
\begin{equation}\label{113}
\delta n_1= \delta n_2=  \frac{ V_0}{\pi^2\gamma\xi^2} {\rm Re}
\int_{-\pi/2}^{\pi/2} d\eta \int_0^{\infty} k d k \frac{  e^{i
{\bf  k r}}}{k^2 - k_+^2} ,
\end{equation}
where $k_+=2\sqrt{s^2 \cos^2 \eta -c_+^2}$. One can  see from Eq.
(\ref{113}) that, first, $\delta n_1=\delta n_2$ and the relative
density remains unperturbed, and, second, the stationary waves
patterns are caused by the upper mode only and emerges at $s>c_+$.
Such features have clear explanation: the obstacle with $V_1=V_2$
cannot excite the mode with $\delta n_1=-\delta n_2$. We emphasize
that these features are specific for the symmetric two-component
condensate. If the bare modes of the components differ from each
other the lower mode is not pure relative density oscillations
($\delta n_1\ne-\delta n_2$) and the obstacle with $V_1=V_2$ may
excite the lower as well as the upper mode.

\begin{figure}
    \includegraphics[width=15cm]{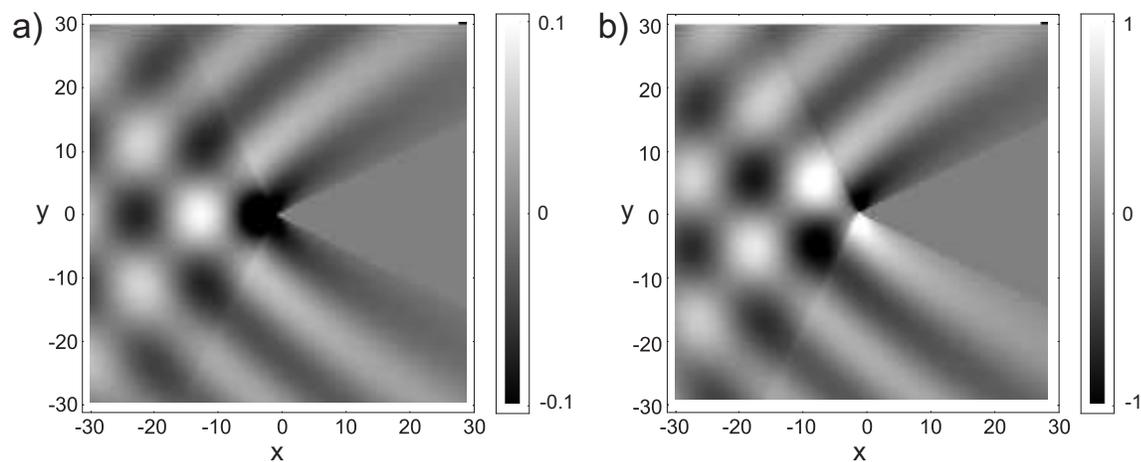}
     \caption {The total density (a) and the relative density (b) patterns (in
relative units) for the stationary waves at $ \theta = \pi/2$ and
$s_1 = s_2 = 0.9$.} \label{f9}
\end{figure}

\begin{figure}
    \includegraphics[width=15cm]{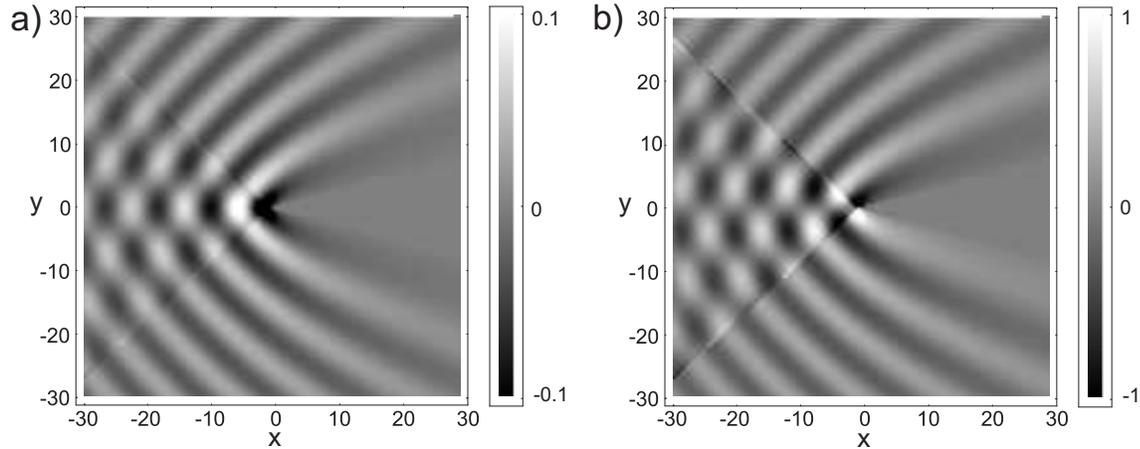}
     \caption {The same as in Fig. \ref{f9}   at $ \theta = 0.45\pi$
and $s_1 = s_2 = 1$.} \label{f10}
\end{figure}

Using Eq. (\ref{112}) and explicit expressions for $C_{i,\lambda}$
and $\kappa_{i,\lambda}$ we obtain the density plots  for the
interfered stationary waves given in Figs. \ref{waves4},
\ref{waves11}. The results  for the total density and relative
density patterns are shown in Figs. \ref{f9}, \ref{f10}.  One can
see that the images presented have clear interference structure
and both the total density and relative density measurements can
be used for the visualization of the stationary waves. Note that
in the case considered the relative density pattern in much more
contrast than the total density pattern.

Ending this section we would note the following. The excitation of
stationary and propagating waves by the obstacle can be considered
as a kind of Cherenkov radiation. Cherenkov radiation arises when
the velocity of a radiating object in the medium exceeds the phase
velocity of radiated waves in this medium. In this respect it was
unclear why  the critical velocity may be larger than the velocity
of the lower mode $c_-$. The answer is the following. At relative
motion of the components the eigenvector for the lower mode
differs from one for the condensate at rest or at $s_-=0$. In
other words, in the flow with $s_- \neq 0$ the structure of the
modes is modified, and the mode radiated at nonzero relative
velocity is not the $c_-$ lower Bogolyubov mode.

\section{Conclusions}
\label{sec5}

In conclusion, the properties of stationary waves arising in a
flow of a two-component BEC past an obstacle have been studied.
The problem was considered for a special case of the symmetric
two-component system (with the components of equal masses of the
particles, equal densities and equal interaction constants).
Nevertheless, the majority of conclusions  are applicable also to
the cases in which such a symmetry between the components is
absent. Let us recite these conclusions.

In a two-component flowing superfluid system the energies of the
excitations can take on not only negative, but complex values. At
reaching of negative values the Landau criterium is broken, and at
reaching of complex values the system becomes unstable with
respect to a spatial separation. The Landau criterium is a joint
condition of both superfluid velocities (and the angle between
them) in the lab reference frame. The stability condition with
respect to a spatial separation is a condition solely on an
absolute value of the relative velocity (which does not depend on
the reference frame). In case of equal (in modulus and in
direction) velocities of the components a spatial separation does
not arise at any velocities. Under increase of the velocities the
Landau criterion is broken first, and then the stability condition
is broken. Stationary waves arise when the Landau criterion is
already broken, but the system remains stable with respect to a
spatial separation. At equal velocities the stationary waves
generated by the lower and the upper modes can arise. If only one
component flows, the  one set of stationary waves (corresponding
to the lower mode) may emerge. If the angle between the velocities
is close to $ \pi/2$, two sets of stationary waves can arise, and
both of them correspond to the lower mode. In general, the
stationary waves are visible at the total density and relative
density images.

We did not consider here the ways of creation of relative flow of
the components in two-component atomic vapors, but we would like
to mention some other possibilities. It is quite simple to realize
such a flow for two components separated with a thin barrier. In
this case the interaction between the components should contain a
long-range (for example, dipole) part. Strictly speaking, in such
systems the spectrum of excitations may differ from (\ref{42})
(due to a long-range interaction). Nevertheless one can expect
that the stationary waves will behave qualitatively the same as in
the case considered in this paper. Similar phenomena may emerge in
some other systems, for example, multilayer electronic systems
with superfluid indirect excitons \cite{kf}. Another perspective
object is superfluid polaritons in semiconductor microcavities,
where superflow can be controlled by the laser beam. In a
one-component polariton system the observation of stationary waves
was reported recently \cite{amo}. Relative motion of the
components may also arise if one component is electrically charged
and the system is subjected by an electromagnetic field. For
instance, such a situation takes place in neutron stars
\cite{bab}.

%


\end{document}